\documentclass[fleqn,twoside]{article}
\usepackage{espcrc2}

\usepackage{graphicx}
\usepackage[figuresright]{rotating}

\newcommand{\AmS}{{\protect\the\textfont2
  A\kern-.1667em\lower.5ex\hbox{M}\kern-.125emS}}

\title{
   \vspace*{-35pt}
   {\normalsize \hfill {\sf KEK-CP-131}} \\
   \vspace*{20pt}
   Light hadron spectrum with two flavors of $O(a)$ improved 
   dynamical quarks : 
   final results from JLQCD\thanks{Talk presented by T.~Kaneko}
}

\author{
   JLQCD Collaboration:
   T.~Kaneko\address[KEK]{High Energy Accelerator Research Organization
                          (KEK), Tsukuba, Ibaraki 305-0801, Japan},
   S.~Aoki\address[Tsukuba]{Institute of Physics, University of 
                            Tsukuba, Tsukuba, Ibaraki 305-8571, Japan},
   R.~Burkhalter$^{\rm b,}$\address[RCCP]{Center for Computational Physics, 
                               University of Tsukuba, Tsukuba, 
                               Ibaraki 305-8577, Japan},
   M.~Fukugita\address[ICRR]{Institute for Cosmic Ray Research, 
                             University of Tokyo, 
                             Kashiwa 277-8582, Japan},
   S.~Hashimoto\addressmark[KEK]
   K-I.~Ishikawa\addressmark[RCCP],
   N.~Ishizuka$^{\rm b,c}$, 
   Y.~Iwasaki$^{\rm b,c}$, 
   K.~Kanaya$^{\rm b,c}$, 
   Y.~Kuramashi\addressmark[KEK], 
   M.~Okawa\address[Hiroshima]{Department of Physics, Hiroshima 
                               University, Higashi-Hiroshima, Hiroshima
                               739-8526, Japan}
   N.~Tsutsui\addressmark[KEK], 
   A.~Ukawa$^{\rm b,c}$, 
   N.~Yamada\addressmark[KEK], 
   T.~Yoshi\'e$^{\rm b,c}$ 
}

\newcommand{\be}{\begin{equation}}
\newcommand{\ee}{\end{equation}}
\newcommand{\bea}{\begin{eqnarray}}
\newcommand{\eea}{\end{eqnarray}}
\newcommand{\bi}{\begin{itemize}}
\newcommand{\ei}{\end{itemize}}
\newcommand{\ben}{\begin{enumerate}}
\newcommand{\een}{\end{enumerate}}
\newcommand{\bt}{\begin{tabbing}}
\newcommand{\et}{\end{tabbing}}

\begin{document}

\begin{abstract}
We present the final results of the JLQCD calculation of 
the light hadron spectrum and quark masses 
with two flavors of dynamical quarks 
using the plaquette gauge action and 
fully $O(a)$-improved Wilson quark action at $\beta=5.2$.
We observe that sea quark effects lead to 
a closer agreement of the strange meson and baryon masses 
with experiment and a reduction of quark masses 
by 20\,--\,30\,\%.
\end{abstract}

\maketitle


\section{Introduction}

Precise determination of the light hadron spectrum
with dynamical up, down and strange quarks 
is a key step for validating QCD at low energies.
As a step toward this goal,
we have pursued a systematic calculation in two-flavor QCD 
using the non-perturbatively $O(a)$-improved Wilson quark action \cite{Spectrum.Nf2.JLQCD.report}.
In this article, we present the final results of 
the hadron spectrum and quark masses from this calculation.
A test of the one-loop chiral perturbation theory formula
in our lattice data is discussed 
in a separate talk \cite{ChPT_test.Nf2.JLQCD}.

Our simulations were performed at $\beta\!=\!5.2$,
where the scale is $a\!=\!0.0887(11)$~fm,
on $12^3\!\times\!48$, $16^3\!\times\!48$ and $20^3\!\times\!48$ 
lattices.
We simulated five sea quark masses
$K_{\rm sea}\!=\!0.1340$, 0.1343, 0.1346, 0.1350 and 0.1355,
which cover the range of 
$m_{\rm PS,sea}/m_{\rm V,sea}\!=\!0.80$\,--\,0.60.
For further details of the simulations,
we refer the reader 
to the previous reports \cite{Spectrum.Nf2.JLQCD.report}
and a forthcoming paper \cite{Spectrum.Nf2.JLQCD}.
The main progress after the last conference is that 
we accumulated doubled statistics (12000 HMC trajectories)
on the largest lattice.

\section{Finite size effects}


Finite size effects (FSE)
are more pronounced in full QCD 
than in quenched QCD.
Since our largest volume size $\simeq \! (\mbox{1.8~fm})^3$
is not so large, it is important to check FSE in our data.

\begin{figure}[t]
   \vspace{-9mm}
   \includegraphics[width=65mm]{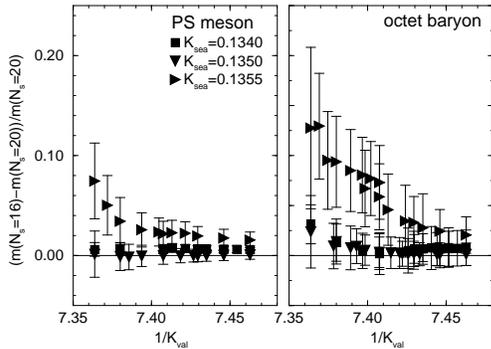}
   \vspace{-12mm}
   \caption
   {
      Relative mass difference between $V\!=\!16^3$ and $20^3$.
   }
   \label{fig:FSE}
   \vspace{-7mm}
\end{figure}

In Fig.~\ref{fig:FSE}, 
we plot the relative mass shift between the two larger lattices
$\Delta m \!=\! (m(N_s\!=\!16)-m(N_s\!=\!20))/m(N_s\!=\!20)$ 
for pseudo-scalar mesons and the octet baryons as a function of 
valence $1/K_{\rm val}$ for each sea quark mass.
We see that 
$\Delta m$ is consistent with zero for the whole range of valence quark mass
down to the second lightest sea quark at $K_{\rm sea}\!\leq\!0.1350$.
The size effect becomes visible and increases with reduced 
$1/K_{\rm val}$ for the 
lightest sea quark at $K_{\rm sea}\!=\!0.1355$.  

The wrapping of valence quarks in the spatial directions 
leading to FSE due to the size of hadrons is suppressed 
by the center $Z(3)$ symmetry in quenched QCD \cite{FSE}.
In full QCD, $Z(3)$ symmetry is broken by the wrapping of sea quarks in the 
spatial directions, whose magnitude increases toward lighter sea quark. 
A qualitative understanding of the behavior in Fig.~\ref{fig:FSE} would be 
that the $Z(3)$ breaking turns on rather quickly around our lightest sea 
quark. 

The volume dependence of our data is well
described by the power law ansatz
$m(N_s) - m(N_s\!=\!\infty) \propto 1/N_s^3$ \cite{FSE}. 
For pseudo-scalar and vector mesons,   
the magnitude of FSE indicated by this fit is about 3~\% 
at $K_{\rm val}\!=\!0.1350$,
which roughly corresponds to the strange quark mass,
and increases to 5\,\% toward the chiral limit. 
However, since the volume dependence is expected to turn into
a milder behavior $\exp [ - m_{\rm PS} N_s]$
for sufficiently large volumes, 
the actual size of FSE should be smaller.  
In particular, at strange quark mass, FSE should not exceed a few \% level. 

Finite size effects are more pronounced for baryon masses. 
For our lightest sea quark, the magnitude of FSE is 
about 5\,\% at $K_{\rm val}\!=\!0.1350$ 
and increases to 10\,\% 
for $K_{\rm val}\!=\!K_{\rm sea}\!=\!0.1355$.
This is comparable with the typical size of the quenching error.
Therefore, masses of light baryons such as $m_N$ and $m_{\Delta}$ 
may be affected by FSE.

\vspace*{-1mm}
\section{Hadron masses} 

\begin{figure}[t]
   \vspace{-9mm}
   \includegraphics[width=63mm]{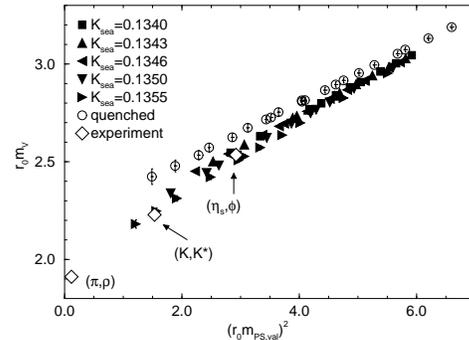}
   \vspace{-12mm}
   \caption
   {
      Vector meson mass as a function of pseudo-scalar meson 
      mass squared.
   }
   \label{fig:mV_vs_mPS2}
   \vspace{-9mm}
\end{figure}

In Fig.~\ref{fig:mV_vs_mPS2},
we compare the valence quark mass dependence of the vector meson 
mass at each sea quark mass and in quenched QCD.
The full QCD data have a steeper slope than the quenched QCD, 
and show a better agreement with the experimental points.

\begin{figure}[t]
   \vspace{3mm}
   \includegraphics[width=63mm]{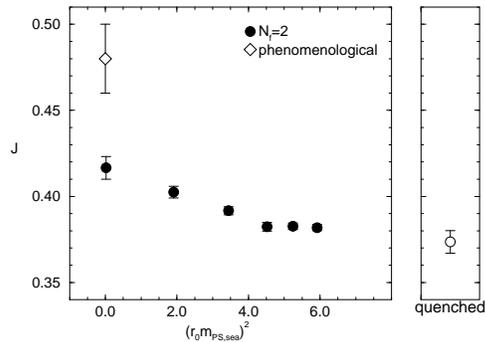}
   \vspace{-12mm}
   \caption
   {
      $J$ parameter in full (left panel) and quenched QCD
      (right panel).
   }
   \label{fig:J}
   \vspace{-7mm}
\end{figure}

The increase of the slope can be explicitly seen 
in the $J$ parameter \cite{J} shown in Fig.~\ref{fig:J}.
The value of $J$, which is consistent with the quenched value 
for heavier sea quark masses,
increases as the sea quark mass decreases.
The chiral limit is still significantly lower than the 
phenomenological value,
which may be attributed to quenching of strange quarks.

Finite size effects, possibly present at a few percent level, 
do not alter these results since masses at larger volumes 
would show a steeper slope in Fig.~\ref{fig:mV_vs_mPS2}.

The hadron spectrum at the physical quark mass
is calculated using data on the largest lattice.
%
%
%
%
We use $m_{\pi}$ and $m_{\rho}$ to fix 
the scale and the light quark mass $m_{ud}$.
The strange quark mass $m_s$ is determined from 
either $K$ or $\phi$ meson mass.

\begin{figure}[t]
   \vspace{-9mm}
   \hspace{5mm}
   \includegraphics[width=65mm]{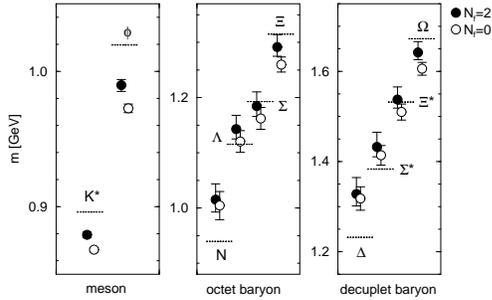}
   \vspace{-10mm}
   \caption
   {
      Hadron spectrum with $K$-input.
   }
   \label{fig:spectrum}
   \vspace{-7mm}
\end{figure}

Our results for the spectrum are plotted in Fig.~\ref{fig:spectrum}.
The deviation of the meson spectrum from experiment 
is reduced significantly by the inclusion of dynamical up and down 
quarks. 
The strange baryon masses, such as $m_{\Xi}$ and $m_{\Omega}$,
also show a closer agreement with experiment in full QCD.
However, sea quark effects are less clear in lighter baryons
like $m_N$ and $m_{\Delta}$.
This is probably due to FSE, which is enhanced for the lighter baryons.

We make a comparison of $m_{K^*}$ with $K$-input
with the CP-PACS results \cite{Spectrum.Nf2.CP-PACS}
in Fig.~\ref{fig:scaling}.
A good agreement with their result in the continuum limit
suggests that the scaling violation in our data is small.

\vspace*{-1mm}
\section{Quark masses}

Calculation of $m_{ud}$ and $m_s$ is performed 
with the bare quark mass defined through either 
the vector (VWI) or axial vector Ward identity (AWI).
For the latter, we use the improved axial current with 
$c_A$ at one-loop \cite{1loop.Zbc}.
The continuum $\overline{\rm MS}$ quark mass is obtained
by the one-loop matching  \cite{1loop.Zbc} 
at scale $\mu\!=\!a^{-1}$
and evolved to $\mu\!=\!2$~GeV using four-loop beta function.

\begin{figure}[t]
   \vspace{-9mm}
   \hspace{5mm}
   \includegraphics[width=70mm]{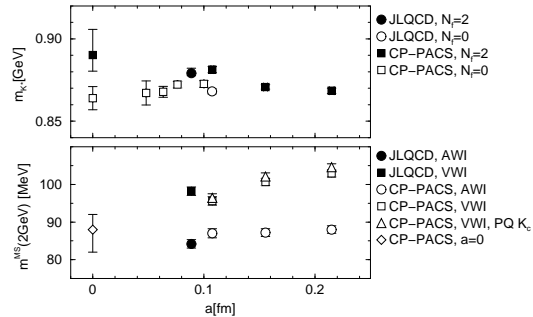}
   \vspace{-16mm}
   \caption
   {
      Comparison of $m_{K^*}$ (top panel) and $m_s$ (bottom panel)
      with CP-PACS results\cite{Spectrum.Nf2.CP-PACS,mq.Nf2.CP-PACS}.
      All results are obtained with $K$-input.
   }
   \label{fig:scaling}
   \vspace{-6mm}
\end{figure}

It is expected that
various systematic uncertainties, such as scaling violation,
partially cancel in the ratio defining the AWI mass 
$m_q\!=\!\langle A_4P^{\dagger} \rangle/(2\langle PP^{\dagger}
\rangle)$.
This is supported by the good agreement with the CP-PACS results
in the continuum limit \cite{Spectrum.Nf2.CP-PACS,mq.Nf2.CP-PACS}
in Fig.~\ref{fig:scaling}.
Such a cancellation is not expected in the VWI mass.  
Indeed there is a sizable difference between our AWI and VWI results, 
and scaling violation is large in the CP-PACS results.


We therefore quote the AWI masses in Table~\ref{tab:mq}
as our central values.
The quark masses decrease by 20\,--\,30\,\%
in two-flavor QCD compared to quenched QCD.
Sea quark effects also reduce 
the difference between $m_s$ with $K$- and $\phi$-inputs.
These are consistent with the observation
in Ref.~\cite{Spectrum.Nf2.CP-PACS,mq.Nf2.CP-PACS}.

\vspace{-1mm}
\section{Conclusions}
\label{sec:conclusion}

We have demonstrated sea quark effects  
at simulation points 
in the meson sector using the $O(a)$-improved Wilson quark action and 
plaquette gauge action at $a^{-1}\!\simeq\!2$~GeV.  
A closer agreement of the strange meson and baryon masses
with experiment is a natural consequence of the finding. 
We also observed that sea quark effects 
lead to a significant reduction of quark masses.

\begin{table}[t]
\caption{
  Results of AWI quark masses in MeV units.
  The input to fix $m_s$ is written in brackets.
}
\begin{tabular}{l|lll}
\hline
$N_f$   & $m_{ud}$  & $ m_s(K)$  & $m_s(\phi)$  \\
\hline
2       & 3.21(4)   & 84.2(1.1)  &  96.3(2.2)   \\
0       & 4.01(8)   & 103.9(1.6) & 127.8(3.3)   \\
\hline
\end{tabular}
\label{tab:mq}
\vspace{-6mm}
\end{table}


\vspace{2mm} 
This work is supported by the Supercomputer Project No.79 (FY2002)
of High Energy Accelerator Research Organization (KEK),
and also in part by the Grant-in-Aid of the Ministry of Education
(Nos. 11640294, 12640253, 12740133, 13135204, 13640259, 13640260,
14046202, 14740173).
N.Y. is supported by the JSPS Research Fellowship.


\end{document}